\documentclass[letterpaper,11pt]{article}

\usepackage{graphicx}
\usepackage{amsmath}
\usepackage{wasysym}
\usepackage{lscape}

\usepackage[T1]{fontenc} 
\usepackage[applemac]{inputenc} 
\usepackage{color}
\usepackage{epstopdf}
\usepackage{setspace}  
\onehalfspace

\begin{document}
\author{Aglaé Kellerer}
\title{Towards wide-field high-resolution retinal imaging }

\maketitle

This is the text of a presentation I gave in 2013 at the conference for {\it Adaptive Optics in Industry and Medicine\/} (AOIM).

\section{Introduction}
We discuss the possibility to apply a novel astronomical technique to retinal imaging. The aim of the method that we suggest is to obtain high-resolution images within large fields-of-view. 

\begin{figure}
\begin{center}
\includegraphics[width=15cm]{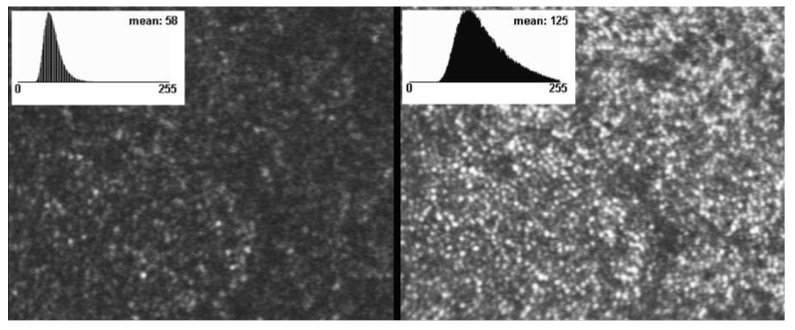}
\caption{Adaptive optics on the eye. Left/right panels: without/with AO correction. From: Roorda et al., Optics Express, 2002. This image has been recorded close to the fovea, where only cone photoreceptors are present. }
\label{fig:1}
\end{center}
\end{figure}

Adaptive optical correction allows one to obtain high-resolution images of the retina. The result is illustrated on Fig.\,\ref{fig:1}: on the left panel, without AO correction, individual cells of the retina are unresolved. On the right panel, after AO correction the individual cones of the retina are distinguishable. These high-resolution images have allowed an improved understanding of the eye’s properties, notably of the photoreceptor distribution. 
One might have expected that the development of AO systems for the retina would lead to a wealth of publications on early-stage diagnoses of retinal diseases. This is however not the case, and we believe that a main reason is the small field size of AO corrected images. 

\begin{figure}
\begin{center}
\includegraphics[width=10cm]{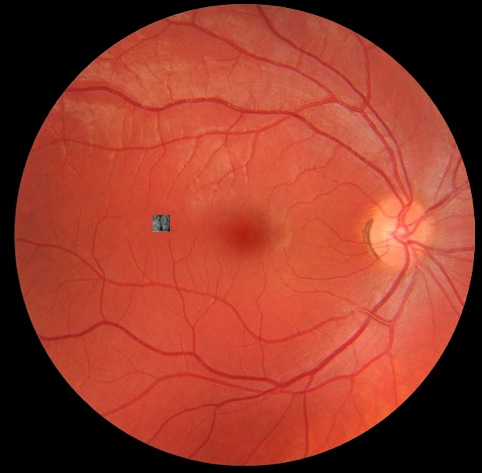}
\caption{ Typical size of an AO corrected image on the retina.}
\label{fig:2}
\end{center}
\end{figure}

Fig.\,\ref{fig:2} shows the typical size of an AO corrected image on the retina. With a conventional AO correction, the resulting image covers a maximum of $3^{\circ}$ on the retina. Several techniques aim at increasing this field of view. One such technique is called star-oriented multi-conjugate approach. It permits to correct images within $6^{\circ}$ fields. Muti-conjugate AO comes in two flavors. Today we will suggest the use of the second flavor – the layer oriented approach. We believe that this shall permit to increase the size of AO corrected images to $10^{\circ}-15^{\circ}$. 

As a bottom line: adaptive optics is a technique that works well on the retina, but it is limited to very narrow fields-of-view. Yet, the diagnosis of many retinal diseases depend upon large field imaging of the retina. 

\section{Current approaches to wide-field, high-resolution imaging}
\begin{figure}
\begin{center}
\includegraphics[width=15cm]{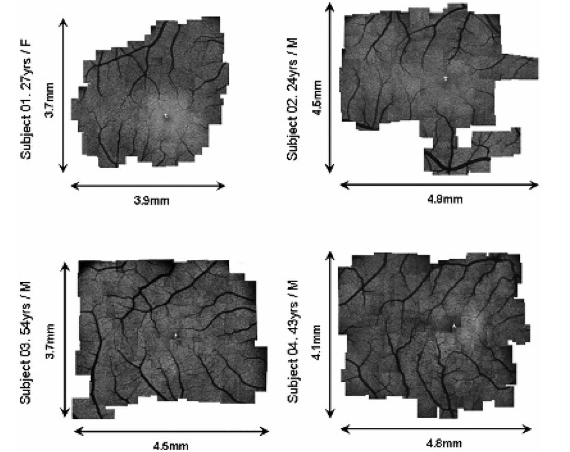}
\caption{One method for wide-field AO imaging relies on the use of a scanning mirror. The individual AO corrected images are small, but a series of images is recorded and are then stitched together to yield one large, high-resolution image. From: Chui et al., JOSA A, 2008}
\label{fig:3}
\end{center}
\end{figure}

One method for wide-field AO imaging relies on the use of a scanning mirror. The individual AO corrected images are small, but a series of images is recorded and are then stitched together to yield one large, high-resolution image. The images in Fig.\,\ref{fig:3} are $10^{\circ}-15^{\circ}$ wide. This method is simple and efficient, but it encounters practical limitations. An imaging session lasts typically one hour and this excludes the examination of young patients. Also, the reconstruction of a large image from a series of small images is not straightforward: ideally, the patient’s visual fixation should be perfectly steady over all acquisitions. Large variations in the fixation lead to artifacts in the image reconstruction. This is a particular difficulty for patients with visual impairment, but variations occur even in healthy eyes, when the subject is trying to maintain a steady fixation. 

Let’s look at the cause of the field-of-view limitations in AO: wavefronts that are focused on different parts of the retina, cross different parts of the eye and thus incur different distortions. An AO correction that is valid along one direction is not valid along another direction. Assume that wavefront distortions are exclusively introduced at the entrance of the optical lens -- this is the eye's pupil, where wavefronts from all directions overlap. The distortions would then be the same whatever the field-direction. In this case the AO correction would be valid within large fields-of-view. In actuality, however, the distortions are introduced in at least 4 different planes: the front and back of the cornea and the front and back of the lens. Additional distortions are introduced by an index gradient within the optical lens. The temporal variations in these aberrations are caused by tear-film motions, blood pulsation and eye-movements. 
In multi-conjugate adaptive optics, deformable mirrors are optically conjugated to the planes where distortions are introduced. Each mirror corrects the distortions of its conjugate plane. 

\begin{figure}[htbp]
\begin{center}
\includegraphics[width=.45\textwidth]{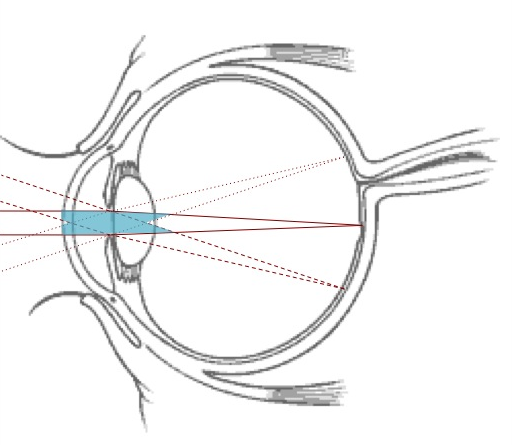}
\includegraphics[width=.45\textwidth]{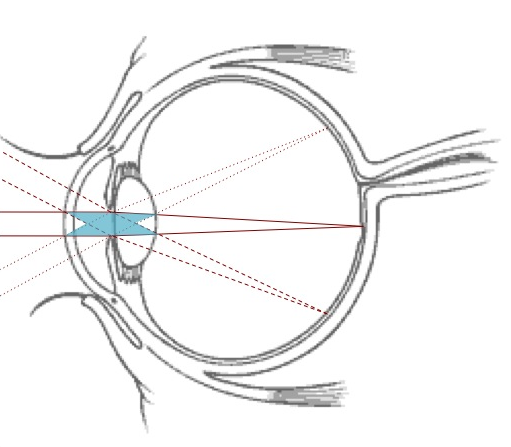}
\caption{Star-oriented multi-conjugate adaptive optics: wavefronts distortions are sensed along several directions. The tomographic reconstruction is effective in those areas where at least two wavefronts overlap: in the blue zone.}
\label{fig:4}
\end{center}
\end{figure}

This requires knowledge of the distribution of optical index variations in the eye: a sensing stage is needed to tomographically reconstruct the profile of wavefront distortions. In a star-oriented multi-conjugate approach, wavefronts distortions are sensed along several directions. The tomographic reconstruction is done computationally and is only effective in those areas where at least two wavefronts overlap -- the blue zone on Fig.\,\ref{fig:4}. Outside that zone, the origin of distortions cannot be determined. 

As the field-size increases, the zone where tomographic reconstruction is effective is reduced (see right panel on Fig.\,\ref{fig:4}) and the quality of the correction therefore decreases. 

\begin{figure}[htbp]
\begin{center}
\includegraphics[width=9cm]{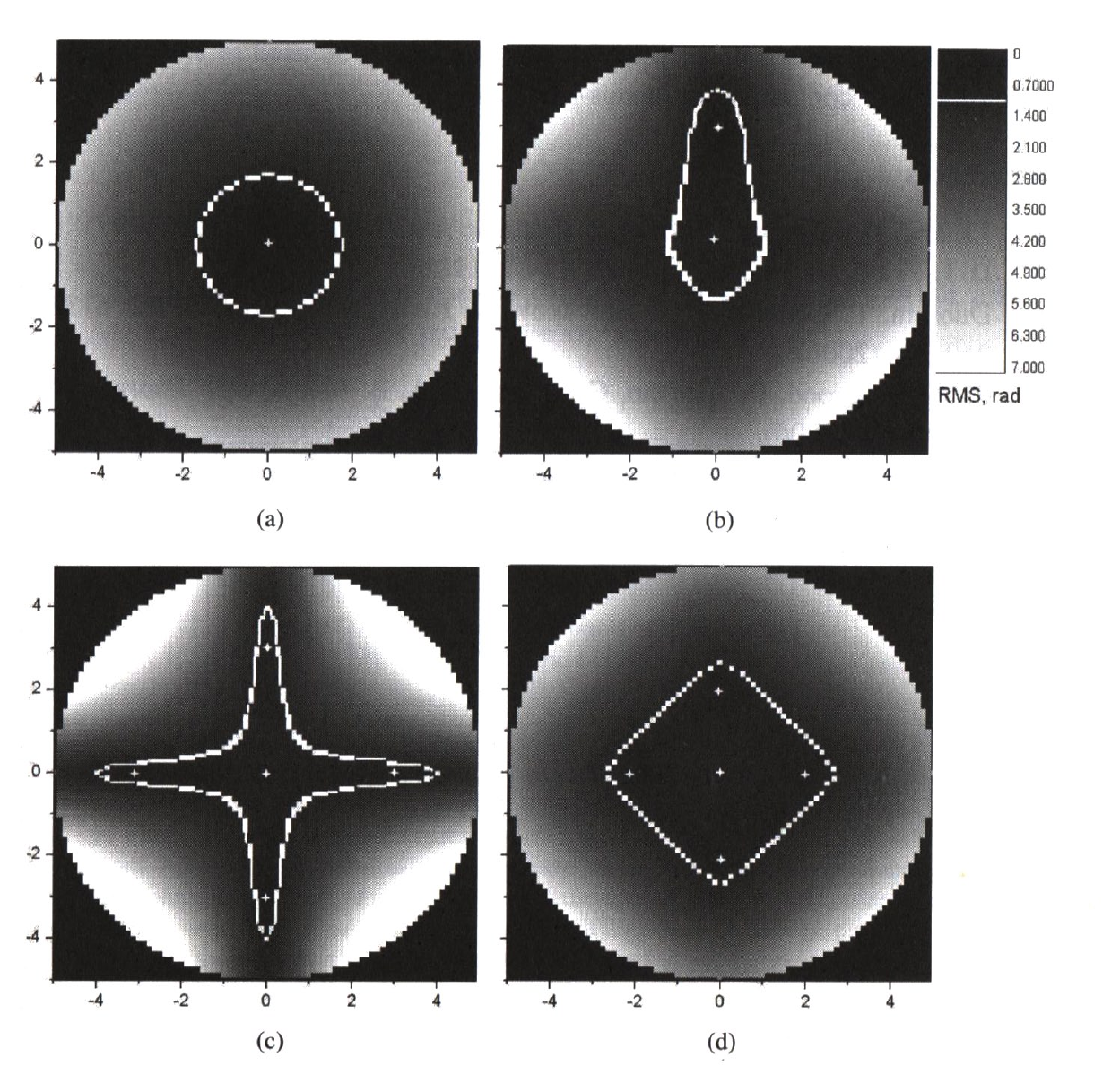}
\caption{Star-oriented multi-conjugate adaptive optics: on fields larger than $\sim 6^\circ$ the tomographic reconstruction becomes inoperable. From: Dubinin et al., proc. of the AOIM conference, 2007}
\label{fig:6}
\end{center}
\end{figure}

The practical consequences were illustrated by Dubinin et al. in a previous AOIM conference: see Fig.\,\ref{fig:6}. The upper left panel shows conventional AO correction -- the correction is good but the field-of-view is fairly limited. The field increases with star-oriented MCAO: see the two panels on the right side. As the field size increases further, the quality of the correction degrades: the correction is then merely valid along the sensing directions, not within the entire field-of-view. 

In conclusion to this section: there are currently two methods to increase the field of AO corrected images: the first involves the use of a scanning mirror. It is simple and efficient, but it suffers practical limitations, notably image artifacts introduced by eye motions and long sessions for the patients. The second method -- the star oriented MCAO -- is limited to $6^\circ$ fields of view: on larger fields the tomographic reconstruction becomes inoperable. 

\begin{figure}[htbp]
\begin{center}
\includegraphics[width=.3\textwidth]{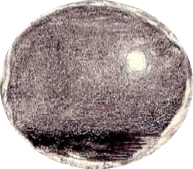}
\caption{An image of the retina -- artist's conception. From: http://www.museumofoptography.net/}
\label{fig:7}
\end{center}
\end{figure}

As an aside: Fig.\,\ref{fig:7} is also an image of the retina, although an artist's conception. It dates back, about one and a half century, to a time when people believed that the last image a person sees before death remains engraved on his retina. Scotland Yard showed interest, as this would be an efficient method to trace a murderer. The victims' corpses were generally discovered too late to find a well preserved retina.  Many rabbits were therefore shown notable scenaries -- such as this seaside at full moon -- before being sacrificed. When this failed to provide observable retinal images, Scotland Yard accepted that it would never see a murderer’s face in a victim’s eye -- all that remained was to see the victim’s future in his palms. But let's return to the more real topic of medical retinal images.

\section{A new approach: layer-oriented multi-conjugate AO}

\begin{figure}[htbp]
\begin{center}
\includegraphics[width=10cm]{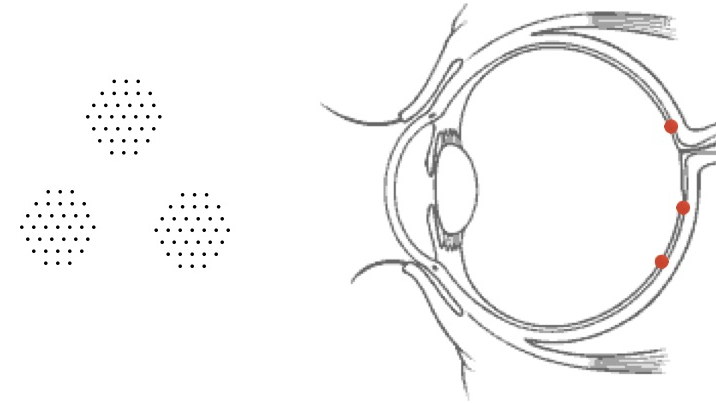}
\caption{Star-oriented multi-conjugate adaptive optics. Several point sources are imaged on the retina: the wavefront distortions are sensed on each of these point sources. Left panel: SH sensor images.}
\label{fig:8}
\end{center}
\end{figure}

In the star-oriented approach to multi-conjugate AO, wavefronts are sensed along several directions: see Fig.\,\ref{fig:8}. To this purpose, several point sources are imaged on the retina: the wavefront distortions are sensed on each of these point sources. 

\begin{figure}[htbp]
\begin{center}
\includegraphics[width=10cm]{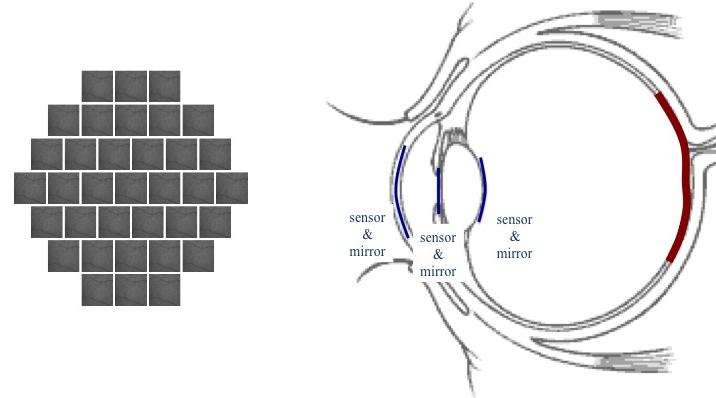}
\caption{Layer-oriented multi-conjugate adaptive optics:  the field of interest is uniformly illuminated and a wide-field SH sensor measures the average wavefront distortions over the field. Left panel: SH sensor images.}
\label{fig:9}
\end{center}
\end{figure}

For a layer-oriented approach, we suggest to uniformly illuminate the field of interest and to sense the average wavefront distortions over the field: see Fig.\,\ref{fig:9}. A sensor is optically conjugated to each of the planes where distortions are introduced. Each sensor averages the distortions over a large field -- close-by turbulence does not vary within the field and is not affected by averaging, distant turbulence varies within the field and is attenuated by the averaging procedure. Each sensor-mirror pair works in a closed loop and corrects the distortions of its conjugate layer. 

Imagine a sensor conjugated to the front of the lens and assume that all distortions are introduced in this same plane: wavefronts from all directions incur the same distortion and each sub-aperture image is therefore globally shifted. 

Assume now that the distortions are introduced at the cornea. Each field direction then incurs different wavefront distortions. As a result the sub-aperture images are distorted rather then globally shifted. 
The measured quantity is the average shift -- it is large if the distortions are introduced in the same plane as the sensor, and small if the distortions are introduced in a distant plane. The measured quantity is thus a good estimate of close-by turbulence: the tomographic reconstruction is done optically and no reconstruction algorithm is needed.

\begin{table*}
\centering
\begin{tabular}{ p{.49\textwidth} p{.49\textwidth}}  \hline
Star-oriented & Layer-oriented \\ \hline
\textcolor{red}{Tomographic reconstruction difficult\/} & \textcolor{blue}{Tomographic reconstruction done optically\/} \\
\textcolor{red}{Limited to fields $<6^\circ$\/} & \textcolor{blue}{Benefits from large fields-of-view\/} \\ 
\textcolor{red}{Correction optimal along direction of guide-stars\/} & \textcolor{blue}{Correction automatically optimized over entire field-of-view\/} \\ 
\textcolor{blue}{No vignetting\/} & \textcolor{red}{Off-pupil sensors are vignetted\/} \\
\textcolor{blue}{Small detectors\/} & \textcolor{red}{Large detectors\/} \\ \hline
\end{tabular}
\caption{ Comparison between star- and layer-oriented multi-conjugate adaptive optics. Advantages in blue, drawbacks in red. }
\label{tab:advantages}
\centering
 \end{table*} 
 
This is the first advantage of the layer-oriented method on the star-oriented approach. Further, as the field-of-view increases the attenuation of distant layers is more efficient: the layer-oriented approach thus benefits from large fields, while the star-oriented approach is limited to $6^\circ$ fields. Finally, the distortions are sensed continuously over the entire field -- thus, the AO loop automatically converges to a correction that is optimal for the entire field. This is in opposition to the star-oriented approach where the distortions are sensed along a few discrete directions. A priori, the loop therefore converges to a correction that is merely optimized along the sensing directions. 
The layer-oriented approach has the following disadvantages: 
 
- Sensors that are not conjugated to the pupil are vignetted. This complicates the estimation of the wavefront slopes: prior to the cross-correlation, the image and the reference need to be multiplied with an appropriate masking function. This is however not a major complication. 

- The major disadvantage of the layer-oriented approach is the need for very high resolution detectors: The detector pixels need to resolve small image shifts, and, each sub-aperture needs to image the entire field of view. 

We first proposed this set-up for solar imaging (Kellerer, Applied Optics, 2012). The technical realization is then very demanding  because daytime turbulence is fast and the detectors must read out at several thousand Hertz. Retinal imaging requires much more moderate correction frequencies of $\sim 30$\,Hz: The detectors, thus, need to be large ($\sim 5000\times 5000$\,pixels) but they do not need to be extremely fast.
 
In conclusion, the eye appears to be well suited for the layer-oriented approach. Of course, extensive experimental testing still needs to be done, and our presentation was meant to generate the necessary interest.

  \end{document}